\documentclass[lettersize,journal,twoside,web]{IEEEtran}
\usepackage{algorithmic}
\usepackage{algorithm}
\usepackage{array}
\usepackage{textcomp}
\usepackage{stfloats}
\usepackage{url}
\usepackage{verbatim}
\usepackage{graphicx}
\usepackage{cite}
\usepackage{amsmath,amssymb,amsfonts}
\usepackage{multirow}
\usepackage{subcaption}

\usepackage[T1]{fontenc}
\usepackage[latin1]{inputenc}

\begin{document}

\title{Fading in reflective and heavily shadowed industrial environments with large arrays}
\author{Sara Willhammar, \IEEEmembership{Graduate student member, IEEE}, Liesbet Van der Perre, \IEEEmembership{Member, IEEE}, and Fredrik Tufvesson, \IEEEmembership{Fellow, IEEE}

\thanks{Manuscript received...}
\thanks{This work has been performed in the framework of the H2020 project 5G-SMART co-funded by the EU under grant number 857008. 
This work has also received funding from ELLIIT.}
\thanks{S. Willhammar, F. Tufvesson and L. Van der Perre are with the Department of Electrical and Information Theory, Lund University, Sweden (e-mail: \{Sara.Gunnarsson, Fredrik.Tufvesson\}@eit.lth.se).}
\thanks{S. Willhammar and L. Van der Perre are also with the Department of Electical Engineering, KU Leuven, Belgium (e-mail: \{Liesbet.Vanderperre\}@kuleuven.be).}}



\maketitle

\begin{abstract}
One of the use cases for 5G systems and beyond is ultra-reliability low-latency communication (URLLC). An enabling technology for URLLC is massive multiple-input multiple-output (MIMO), which can increase reliability due to improved user separation, array gain and the channel hardening effect. Measurements have been performed in an operating factory environment at 3.7~GHz with a co-located massive MIMO array and a unique randomly distributed array. Channel hardening can appear when the number of antennas is increased such that the variations of channel gain (small-scale fading) is decreased and it is here quantified. The cumulative distribution function (CDF) of the channel gains then becomes steeper and its tail is reduced. This CDF is  modeled and the required fading margins are quantified. By deploying a distributed array, the large-scale power variations can also be reduced, further improving reliability. The large array in this rich scattering environment, creates a more reliable channel as it approaches an independent identically distributed (i.i.d.) complex Gaussian channel, indicating that one can rethink the system design in terms of e.g. channel coding and re-transmission strategies, in order to reduce latency. To conclude, massive MIMO is a highly interesting technology for reliable connectivity in reflective and heavily shadowed industrial environments.
\end{abstract}



\section{Introduction}

Wireless cellular communication is constantly evolving; currently 5G is rolled out and research is aiming towards 6G. 
One focus area for 5G systems, for which the requirements still partially are unfulfilled and hence will remain a focus area, is the ultra-reliability low-latency communication (URLLC) use case. Therefore, for 6G systems, discussions are ongoing regarding neXt generation URLLC (xURLLC). It has applications within e.g. automotive industry, remote surgery and smart manufacturing. URLLC has gained huge interest and commonly mentioned goals are reliabilities of $10^{-9}-10^{-5}$ and latencies of 1~ms~\cite{urllc-5g}.

To achieve the goals of xURLLC, new technologies are required. From a physical layer point of view, one of these enabling technologies is massive multiple-input multiple-output (MIMO) \cite{Marzetta2010}.
The deployment of massive MIMO systems enables characteristics that can improve reliability. First of all, due to favorable propagation conditions -- where the channels between two users become pairwise orthogonal -- the possibility to separate spatially multiplexed users is improved. Secondly, when combining the many antennas, the array gain results in a stronger received signal. Lastly, the channel becomes more flat in both time and frequency as a channel hardening effect appears; this means that the variations of channel gain decrease as the number of antennas increases. These two characteristics in combination leads to a decreased probability of outage, which can be so low that the way the rest of the communication system is designed needs to be re-considered. 

Channel hardening has been investigated in theory\cite{Marzetta2010, Bjoernson2017, downlink_pilots, chhard_phy_model, 8902768, proof_conv} where an independent and identically distributed (i.i.d.) complex Gaussian channel model is commonly used. However, in an indoor scenario, the model has been shown to be overly-optimistic~\cite{chhard_journal} since real massive MIMO channels usually are spatially correlated.
Lately, there have been several studies investigating channel hardening in simulations~\cite{chhard_sim},
experimentally~\cite{chhard_aalborg, chhard_aalborg_2, Ghiaasi2019, chhard_journal} and also for systems beyond conventional massive MIMO such as cell-free massive MIMO~\cite{cell_free} and intelligent surfaces~\cite{chhard_ris}.

With the channel hardening effect, the cumulative distribution function (CDF) of channel gains becomes steeper and its tails smaller. 
A number of studies have proposed solutions to model the channel and the tails of these CDFs for URLLC purposes.
As URLLC is about accounting for the extreme outage events that rarely happen, an extreme value theory approach is applicable to model the channel~\cite{evt_chan_mod}. In \cite{urllc_chan_mod} approximations for the tail distribution for a wide range of channel models are presented, consisting of only two parameters: an exponent and an offset. First elaborating on the benefits and challenges when using parametric and non-parametric channel models, \cite{stat_approach} concludes that the power law tail approximation in \cite{urllc_chan_mod} provides a good trade-off. The concept of local diversity, which concerns evaluating the slope of the CDF at a certain point, is presented in \cite{local_div}.
The CDFs can also be used to relate to the required fading margin, where a definition is provided in~\cite{fading_margin}.



\begin{figure*}
    \centering
    \hspace{-0.5in}
    \begin{subfigure}[b]{0.3\textwidth}
        \centering
        \resizebox{2in}{!}{\includegraphics[height=0.3\textwidth]{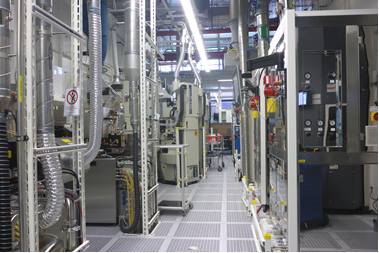}}
        \caption{View towards the co-located array.}
        \label{fig:co}
    \end{subfigure}
    \begin{subfigure}[b]{0.3\textwidth}
        \centering
        \resizebox{2in}{!}{\includegraphics[height=0.3\textwidth]{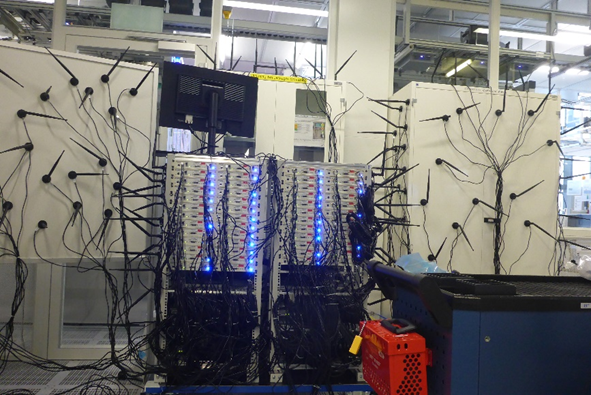}}
        \caption{The distributed array.}
        \label{fig:dist}
    \end{subfigure}
    \begin{subfigure}[b]{0.3\textwidth}
        \centering
        \resizebox{2.38in}{!}{\includegraphics[height=0.3\textwidth]{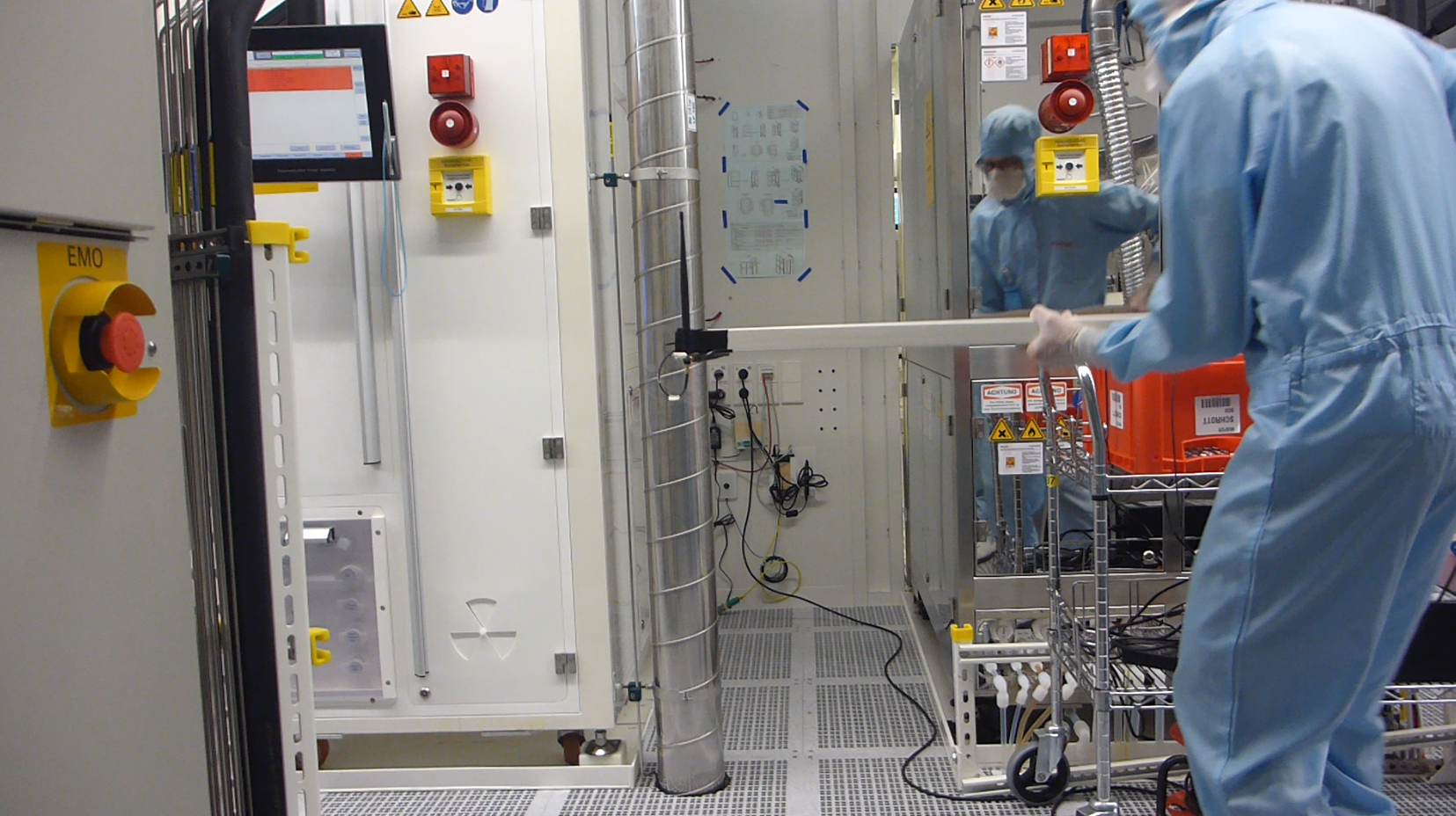}}
        \caption{View towards the left aisle.}
        \label{fig:left}
    \end{subfigure}
    \caption{Pictures from the measured environment in the Bosch semiconductor factory.}
    \label{fig:factory-pics}
    \vspace{-0.3in}
\end{figure*}

With the small-scale variations significantly reduced due to channel hardening, the reliability becomes mostly related to the large-scale power variations, mainly caused by shadowing. To combat shadowing, a distributed setup can be the solution, resulting in that not all antennas are blocked simultaneously. Various distributed configurations (ranging from semi-distributed to fully distributed) have been investigated in different indoor scenarios \cite{dist-3,dist-4,dist-5,dist-6} and their performance has been compared to the co-located equivalent~\cite{dist-3,dist-4}. 

This study contributes to the area by analysing and modeling the channel based on a unique measurement campaign performed in one of Bosch's operating semiconductor factories in Reutlingen, Germany. The factory is an extremely complex environment from a wireless communication perspective.
Channels have been collected with a massive MIMO system equipped with both a co-located and a randomly, fully distributed array (within practical constraints). For the two arrays, the channel hardening effect and required fading margins are quantified. The tails of the channel gain CDFs are characterized. Furthermore, an assessment of the decreased large-scale power variations when deploying a distributed setup is made. To the best of our knowledge, no measurement-based study has been performed with a fully distributed massive MIMO setup in these extreme environments. Our experiments in combination with the analysis, characterization and modeling provide valuable insights to enable URLLC in complex factory environments. 

The structure of the paper is as follows. The measurement scenario in Section~II, followed by measurement setup and data processing are described in Section~II~I. Following these sections, results are presented in Section~IV where a high-level inspection is presented, Section~V concerns the channel hardening effect and the tails of the CDFs are modeled in Section~VI with related fading margins in Section~VII. After the elaboration on small-scale fading, the shadowing-related results are presented in Section~VIII. Lastly, in Section~IX, summary and conclusions are provided.

\vspace{-0.05in}

\section{Measurement scenario}

The measurements have been performed in the Bosch semiconductor factory in Reutlingen, Germany. This factory provides an environment with many walls and long corridors as well as a lot of high metallic machinery, all providing substantial shadowing and reflections of signals, as can be seen in Fig.~\ref{fig:factory-pics}. The above leads to rich scattering. With robots and humans moving around it also makes it a dynamic environment. The scenario investigated here consists of a corridor with tall machinery on the sides. This environment includes many tricky places in terms of coverage, large-scale and small-scale fading.

The scenario is described in Fig.~\ref{fig:factory-pics} where Fig.~\ref{fig:co} shows the view from the corridor towards the base station (BS) with the co-located array deployed. Directly in front of the array there is an open space and all antennas have their highest gain in the direction of where the user equipment (UE) is located. Fig.~\ref{fig:dist} shows the deployed distributed array. Note that this random deployment also means that antennas may have other equipment and/or antennas in the close vicinity and most antennas do not have their highest gain in direction towards the place where the UE is located, which is different from the situation of the co-located array. Lastly, in Fig.~\ref{fig:left}, the view from the corridor towards the left aisle in the corridor can be seen. 

\begin{figure}[h]
\centering
    \begin{subfigure}[b]{0.2\textwidth}
        \centering
        \resizebox{0.7in}{!}{\includegraphics[height=0.1\textwidth]{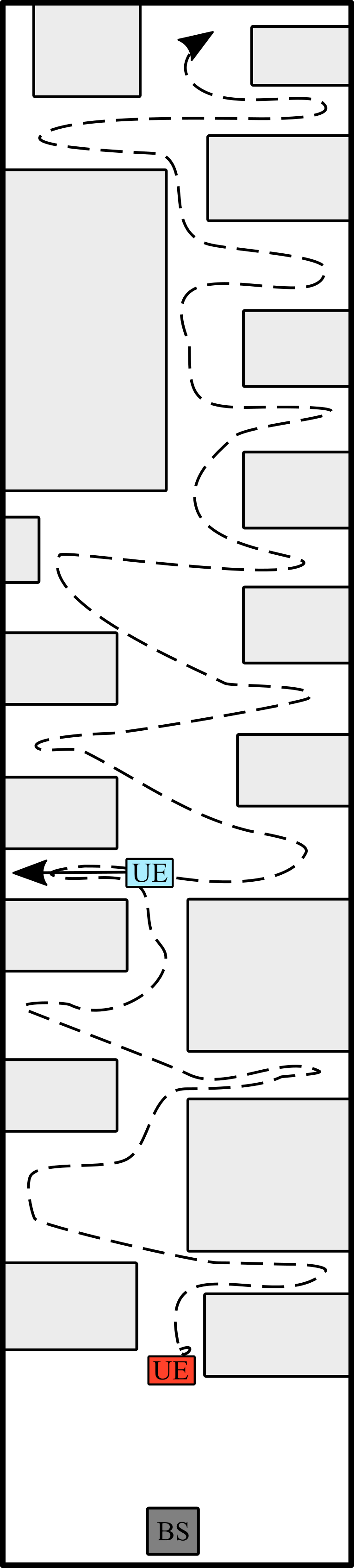}}
        \caption{ }
        \label{fig:map}
    \end{subfigure}
    \begin{subfigure}[b]{0.2\textwidth}
        \centering
        \resizebox{1.6in}{!}{\includegraphics[height=0.5\textwidth]{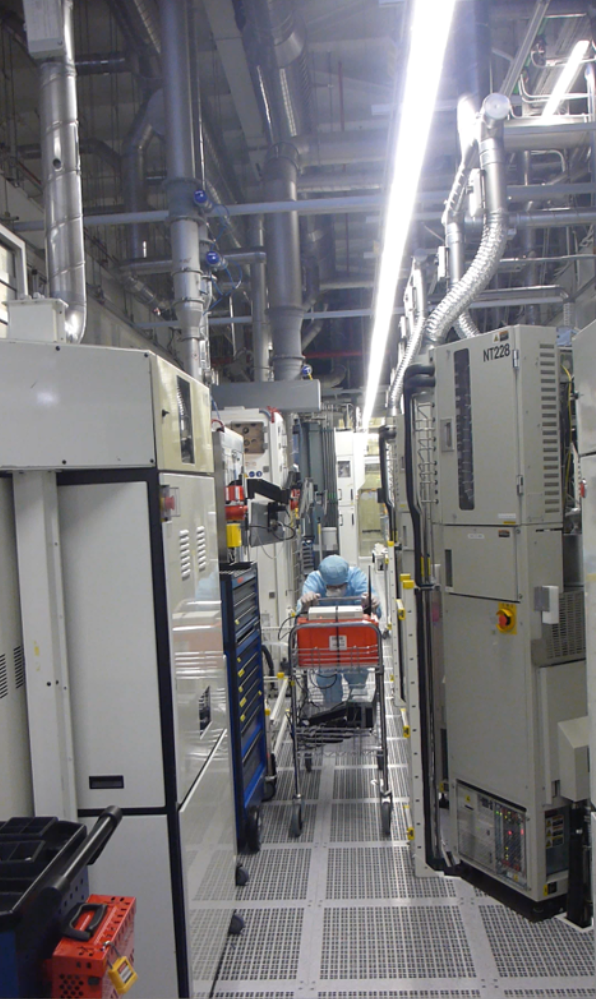}}
        \caption{ }
        \label{fig:corridor}
    \end{subfigure}
    \caption{Overview of the measurement scenario with (a) the map, and (b) the view from the BS towards the UE.}
    \label{fig:map_pics}
    \vspace{-0.2in}
\end{figure}

An overview of where the measurements took place is visualized in Fig.~\ref{fig:map}, where the BS is placed next to a wall and the UE is in the middle of the corridor (the two UEs representing two different experiments). In the first experiment the blue UE moves from line-of-sight (LoS) to non-line-of-sight (NLoS) to the left (as seen from the BS). The UE antennas are at first pointing as in Fig.~\ref{fig:left}, then when moving into NLoS the stick is rotated $90^{\circ}$ to the right. The measurements also include scanning the entire area behind the machinery with 6000 time samples, maintaining the antenna orientations. This experiment serves as a basis for collecting statistics of the small-scale fading in heavily shadowed places. For the second experiment, the red UE starts close to the BS and then moves along the corridor while entering the aisles behind the machinery, aiming at collecting large-scale fading statistics. In Fig.~\ref{fig:corridor} the view from the BS towards the UE and the corridor is seen.

\section{Measurement setup and data processing}
The measurement device used in the experiment is the Lund university Massive MIMO (LuMaMi) testbed , a long-term-evolution (LTE)-like and software-defined radio (SDR)-based testbed with 100~coherent radio frequency (RF)-chains, able to serve up to twelve UEs simultaneously. ~
The equipment used as UEs consists of one SDR, each having two RF-chains connected to dipole antennas with vertical and horizontal orientation. The two RF-chains are independently processed and can be seen as two separate users.

The testbed, acting as the BS, is normally equipped with a co-located array, where the connected antenna ports form a rectangular array with four rows and 25~dual-polarized $\lambda/2$-spaced antenna elements in each row. One polarization per antenna element is connected, where the polarization is alternating between two consecutive elements. The second array is a randomly distributed array with random directions and polarizations, with the aim to increase the exploited diversity in the environment. The distributed array consists of 100~omni-directional dipole antennas which, via a cable of three or five meters, are connected to the testbed. 

\begin{figure}[t]
    \centering
    \includegraphics[width=0.4\textwidth]{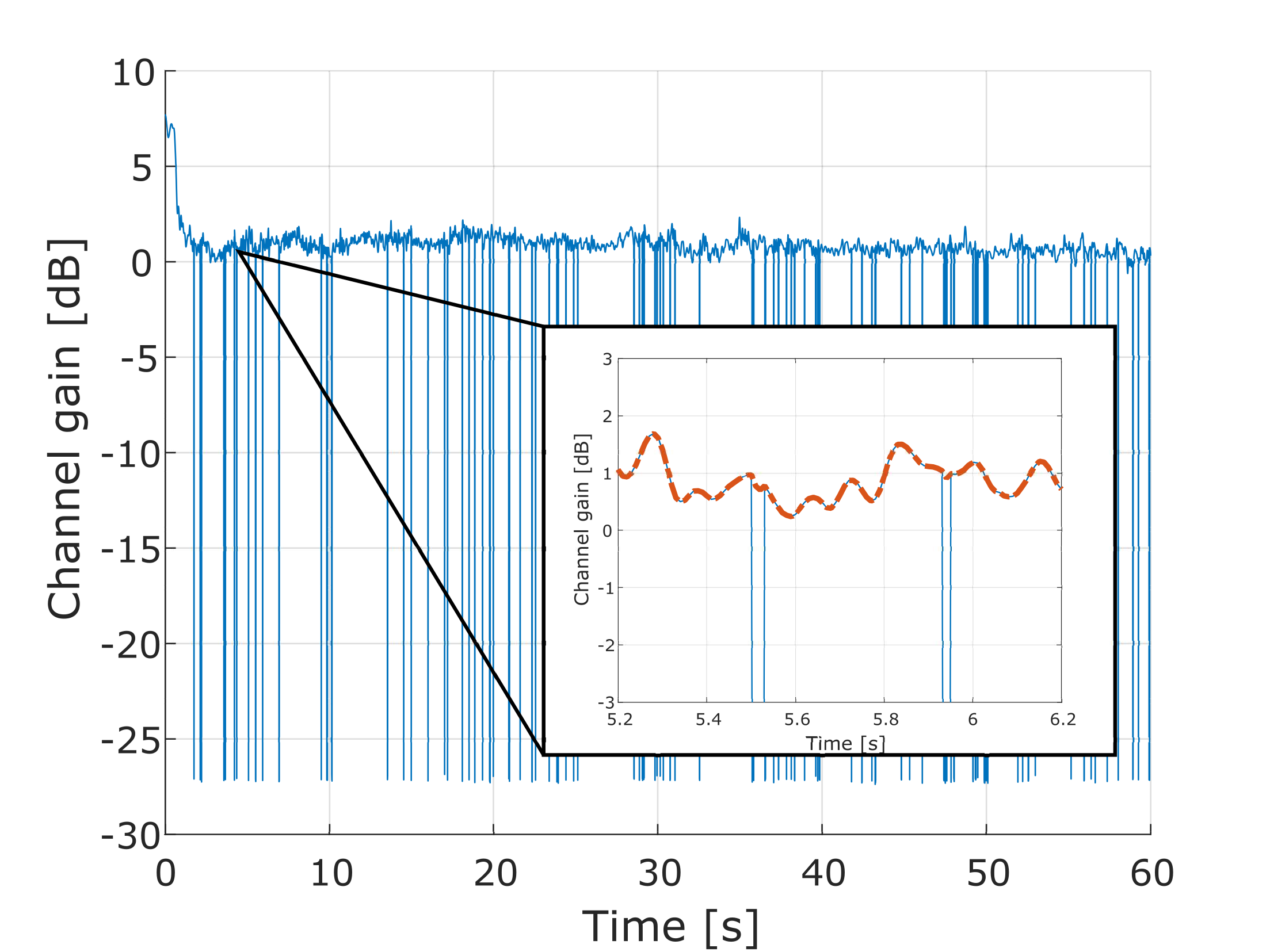}
    \caption{Raw channel gain over time, averaged over frequency and summed for BS antennas, showing lost samples and an example of linear interpolation applied for each frequency point and BS antenna.} 
    \label{fig:raw-signal}
    \vspace{-0.2in}
\end{figure}

The synchronization between the BS and UEs is over-the-air (OTA) and starts when the BS sends out a synchronization signal in the beginning of each frame. When receiving, the UE uses that signal to synchronize to the BS. 
During the measurements, samples have occasionally been lost. This is to our best understanding due to the OTA synchronization failing and hence the corresponding channel sample is not stored\footnote{In other experiments where rubidium clocks have been used as frequency reference, then no samples have been lost. This together with visual inspection of the stored transfer functions point to the synchronization as origin for the lost sample problem.}. 
The raw channel gain $|h|^2$ over time, as averaged over frequency and summed for all BS antennas when using the vertical UE orientation, is shown in Fig.~\ref{fig:raw-signal}. This shows an example of how the lost samples are witnessed in the response, where dips of more than 25~dB can be seen, randomly distributed over time. Most often, only 1-2 samples in a row are lost. Occasionally but quite rarely, 3-4 samples in a row are lost. There are no signs of fading dips when samples are lost.

The first uplink pilot in each frame is logged for later analysis, meaning that a channel sample is collected every 10~ms (the slowest repetition rate that can be handled by the testbed), i.e. $f_{rep}
=100~\text{Hz}$. 
Channels are continuously recorded for the $M=100$~BS antennas at a carrier frequency of 3.7~GHz and over 20~MHz bandwidth, with $F=100$~uniformly distributed frequency samples over the bandwidth. 
In time, $N=6000$~samples are collected. This results in, for each UE, a channel matrix of dimensions $H=[N \times F \times M]$. 

Considering that the repetition frequency needs to be at least twice the maximum Doppler frequency in order to not violate the Nyquist sampling criterion, i.e. $f_{rep}\ge2f_{max}$, the maximum speed ${v_{max}}$ of the UE can be $v_{max}=\frac{cf_{max}}{f_c}$, where $c$ is the speed of light, $f_c$ the carrier frequency and $f_{max}$ is the maximum Doppler frequency that can be handled with this repetition rate. Here this means that the maximum speed of the UE can be 4~m/s; the experiments here are performed below this limit. 

The complex time correlation for all BS antennas and frequency samples was investigated. The results showed that the correlation coefficient was less than 0.5 for samples separated 20~ms or more (i.e. two~samples) for all frequency points and BS antennas. The samples collected are hence indeed quite independent which is good for the statistical analysis but makes it difficult to make a good interpolation. Therefore, when investigating the channel statistics, the outliers due to the OTA synchronization error are removed. When showing the time series, linear interpolation is applied to each frequency point and BS antenna. An example of this is shown in the smaller box in Fig.~\ref{fig:raw-signal} where the original signal with outliers is shown in blue and the interpolated signal in dashed red.

\vspace{-0.1in}


\section{High-level inspection}

\begin{figure}[t]
    \centering
    \includegraphics[width=0.4\textwidth]{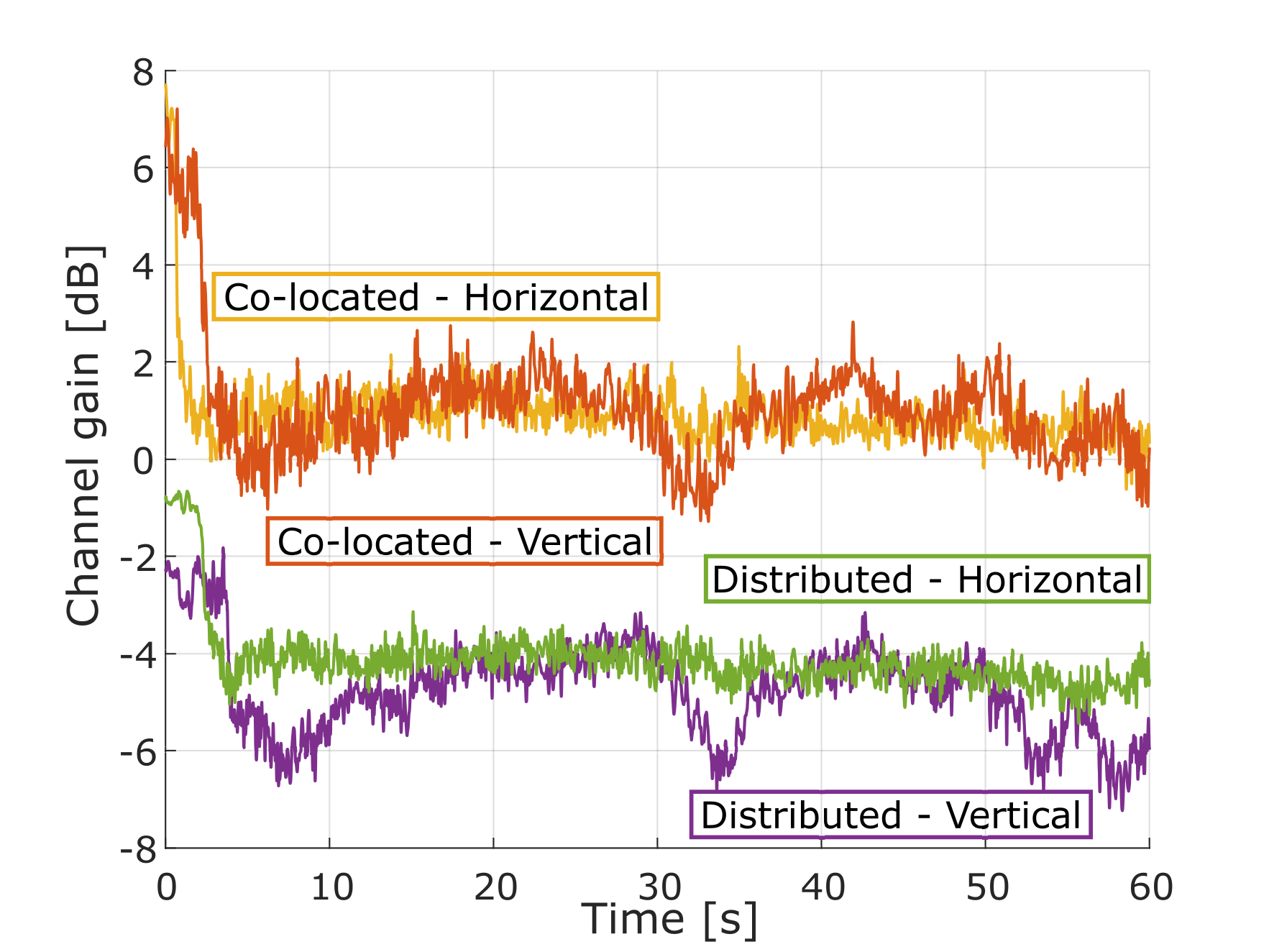}
    \caption{Channel gain for both UE orientations and the co-located and distributed array over time with both UE orientations, when going into the left aisle.
    } 
    \label{fig:ch-gain}
    \vspace{-0.2in}
\end{figure}

Taking a first look at the channel responses from the first experiment -- the UE moving from LoS to NLoS -- Fig.~\ref{fig:ch-gain} shows the interpolated channel gain over time when going into the left aisle for both arrays and UE orientations. The first seconds from the measurement are in LoS, with the higher channel gain, and then the UE is shadowed by the machinery for the rest of the measurements while there scanning the area. The channel gains of the co-located array are higher due to the higher antenna gain and that the distributed setup is also influenced by the attenuation from cables and adaptors. The small dips are when the UE antenna is close to the floor, indicating that coverage can be trickier at these spots.

When being shadowed, the channel gain for the co-located drops about 6~dB and for the distributed setup the gain drops about 2-3~dB, depending on the UE orientation. The LoS can hence be concluded to be less prominent in the distributed case, as expected since the antennas are randomly placed. This leads to somewhat bad placements for this specific measurement, while for the co-located array, all antennas are pointing in the direction where the UE is. In NLoS, the channel gain varies with the movement of the UE, the variations for the vertically oriented antennas are larger than for the horizontally oriented.

\begin{figure}[t]
    \centering
    \includegraphics[width=0.4\textwidth]{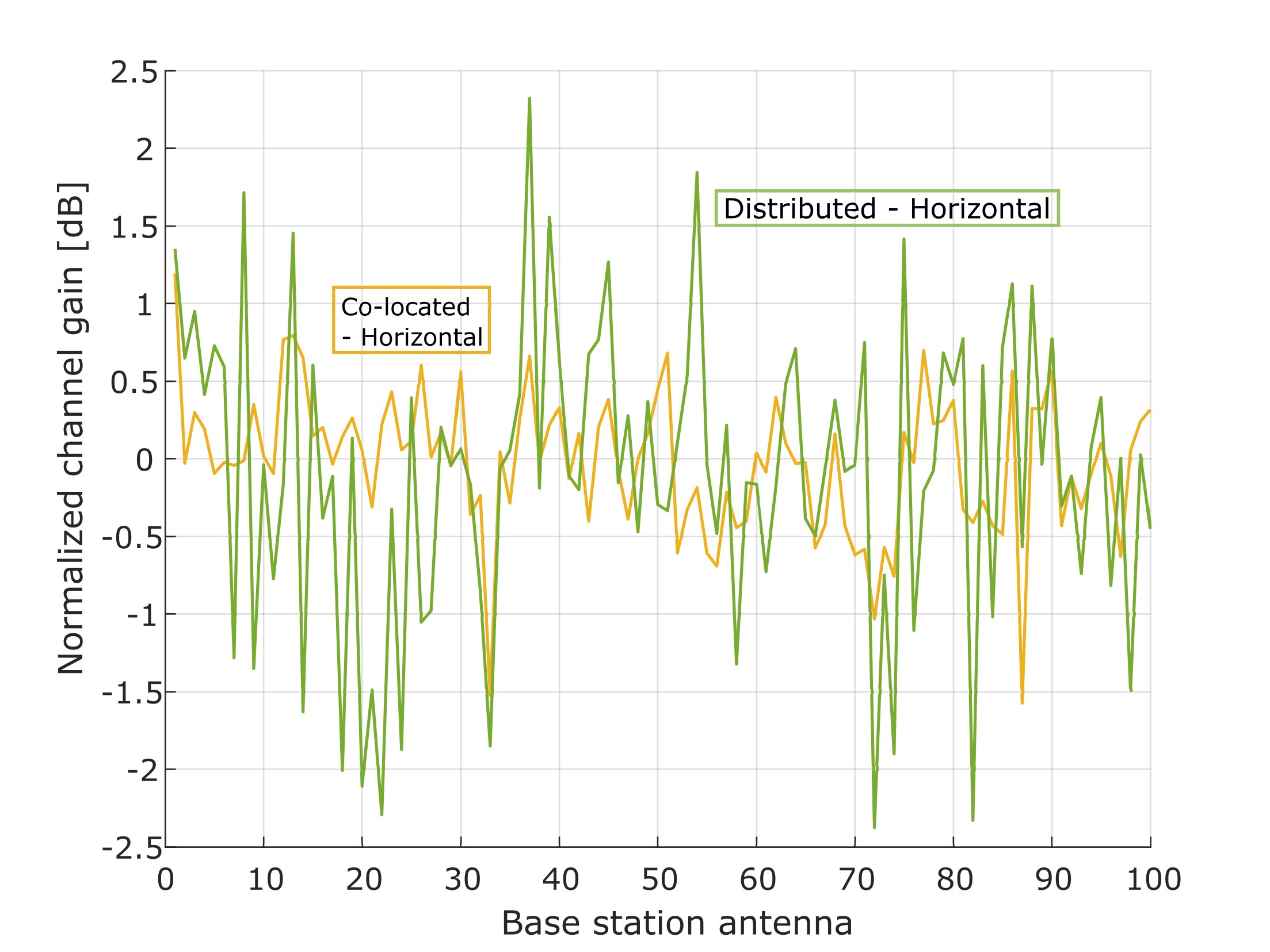}
    \caption{Normalized channel gain, as per BS antenna, in the co-located and distributed array when scanning the left aisle with horizontal UE orientation. 
    }
    \label{fig:mean}
    \vspace{-0.2in}
\end{figure}

For further results and analysis, the first five seconds of each measurement are removed, in order to focus on the heavily shadowed areas. The means and standard deviations of the gain in dB, as starting from five~seconds are summarized in Table.~\ref{tab:mean-std}

\begin{table}[h]
    \vspace{-0.1in}
    \centering
    \begin{tabular}{cc|cc}
        & & $\hat{\mu}$ & $\hat{\sigma}$ \\
        \hline
         \multirow{2}{*}{Co-located} & V & 0.88 & 0.70   \\
         & H & 0.86 & 0.38  \\
         \hline
         \multirow{2}{*}{Distributed} & V & -4.87 & 0.80  \\
         & H & -4.25 & 0.32 \\
    \end{tabular}
    \caption{Estimated parameters starting from 5~seconds in Fig.~\ref{fig:ch-gain} (in dB).}
    \label{tab:mean-std}
    \vspace{-0.2in}
\end{table}

Looking closer into the similarities and differences between the BS antennas of the two arrays, the mean (as averaged over time and frequency) of the normalized channel gain is shown in Fig.~\ref{fig:mean}. It can be seen for the co-located array, most antennas are centered closely around the mean (which is $0$~dB due to the normalization) while for the distributed array, the power variations over the array are larger. For the co-located array the standard deviation is $0.10$~dB and for the distributed array it is $0.22$~dB. The corresponding values for the vertical UE orientation is $0.18$~dB and $0.24$~dB, respectively. This is in line with the assumption that for the distributed case, some antenna will have beneficial placements, while some will contribute very little.

\vspace{-0.1in}

\section{Channel hardening}
Starting the analysis, the channel hardening effect in this environment is investigated and quantified. The definition of channel hardening, as given in \cite{downlink_pilots}, is that the channel vector $\mathbf{h}_{k}$ between the BS and a user $k$ offers channel hardening if

\begin{equation}
    \frac{\text{Var}\{\|\mathbf{h}_{k}\|^2\}}{\text{E}\{\|\mathbf{h}_{k}\|^2\}^2}\rightarrow 0, \hspace{0.5cm} \text{as} \hspace{0.2cm} M\rightarrow \infty,
    \label{eq:chhard_def}
\end{equation}

\noindent where $M$ is the number of base station antennas. The evaluation of channel hardening is performed in the same manner as in \cite{chhard_journal}, and with similarities to the work in \cite{chhard_aalborg,chhard_aalborg_2}. Before analysis, the channel matrix for each UE antenna is normalized as

\vspace{-0.2in}

\begin{equation}
    \mathbf{\overline{h}}_{k}(n,f) = \frac{\mathbf{h}_{k}(n,f)}{\sqrt[]{\frac{1}{N F M}\sum_{n=1}^{N}\sum_{f=1}^F\sum_{m=1}^M|h_{km}(n,f)|^2}},
    \label{eq:norm}
\end{equation}

\noindent for $N$ snapshots, $F$ frequency points and $M$ \textit{selected} BS antennas, where \textit{selected} means those antennas for which the channel hardening is later evaluated. This normalization is also the one used in Fig.~\ref{fig:mean} and means that each entry in $\overline{\mathbf{h}}_{k}$, as averaged over frequency, time, and BS antennas, has an average power equal to one. The channel hardening is then evaluated as the standard deviation (square root of the variance) of channel gain for each UE as






\begin{equation}
    \text{std}_k = \sqrt{\frac{1}{N F} \sum_{n=1}^{N} \sum_{f=1}^{F} |\overline{G}_k(n,f) - \mu_k|^2},
    \label{eq:std}
\end{equation}

\noindent where $\overline{G}_k(n,f)$ is the instantaneous channel gain $|\overline{h}_{km}(n,f)|^2$, as summed and divided by $M$, and $\mu_k$ is the average channel gain, as summed and divided by $N$ and $F$, and hence equal to one (i.e. $0$~dB); details can be found in~\cite{chhard_journal}. With (\ref{eq:std}), the channel hardening can be quantified as the difference in standard deviation when using $1$ and $M$ BS antennas.


\begin{figure}[t]
    \centering
    \includegraphics[width=0.4\textwidth]{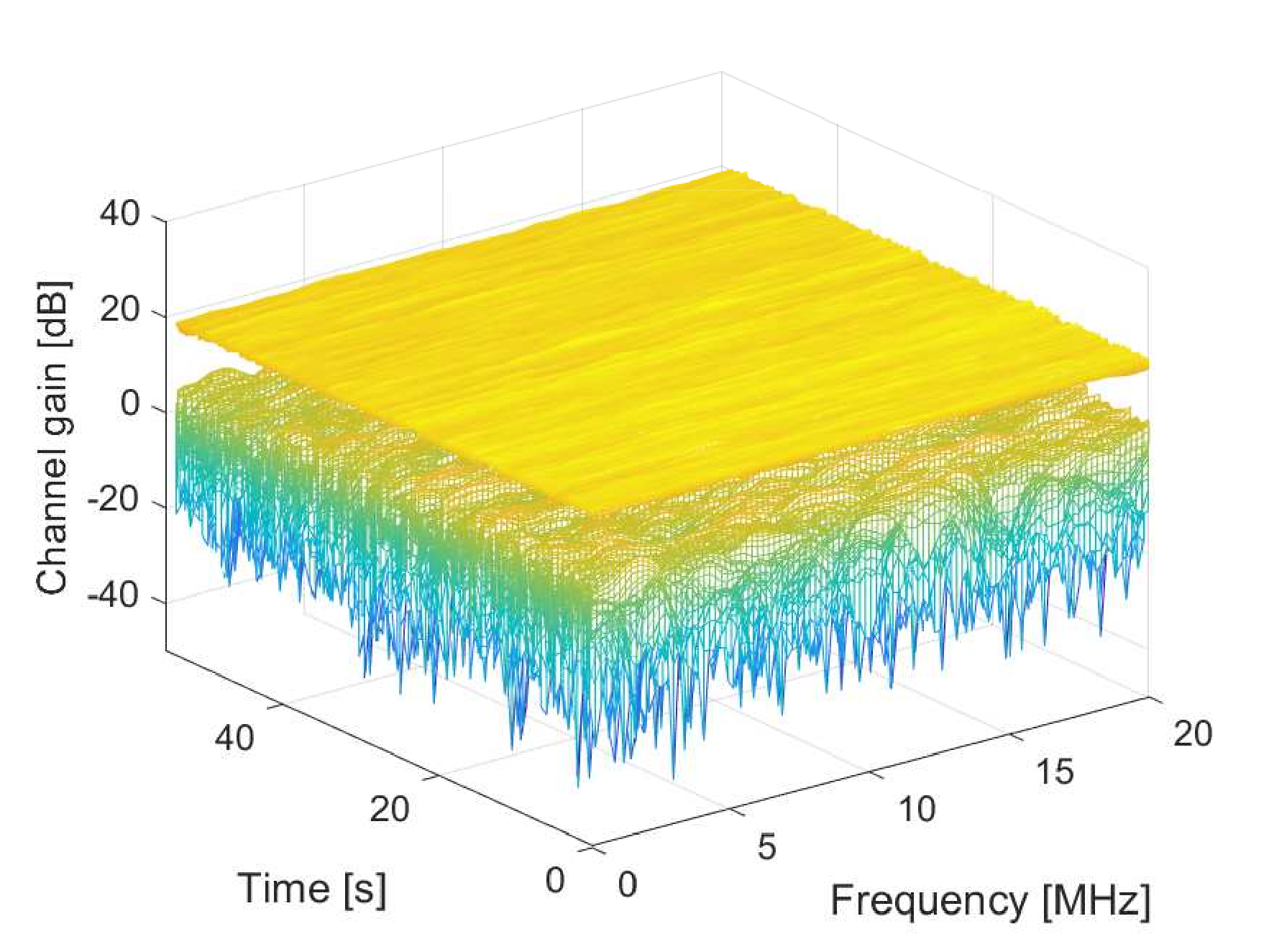}
    \caption{Channel gain over time and frequency for $1$ (lower) and $100$ BS antennas (upper) when going into the left aisle and the co-located array is deployed and horizontal UE orientation.}
    \label{fig:chhard-left}
    \vspace{-0.2in}
\end{figure}

\begin{figure}[t]
    \centering
    \includegraphics[width=0.4\textwidth]{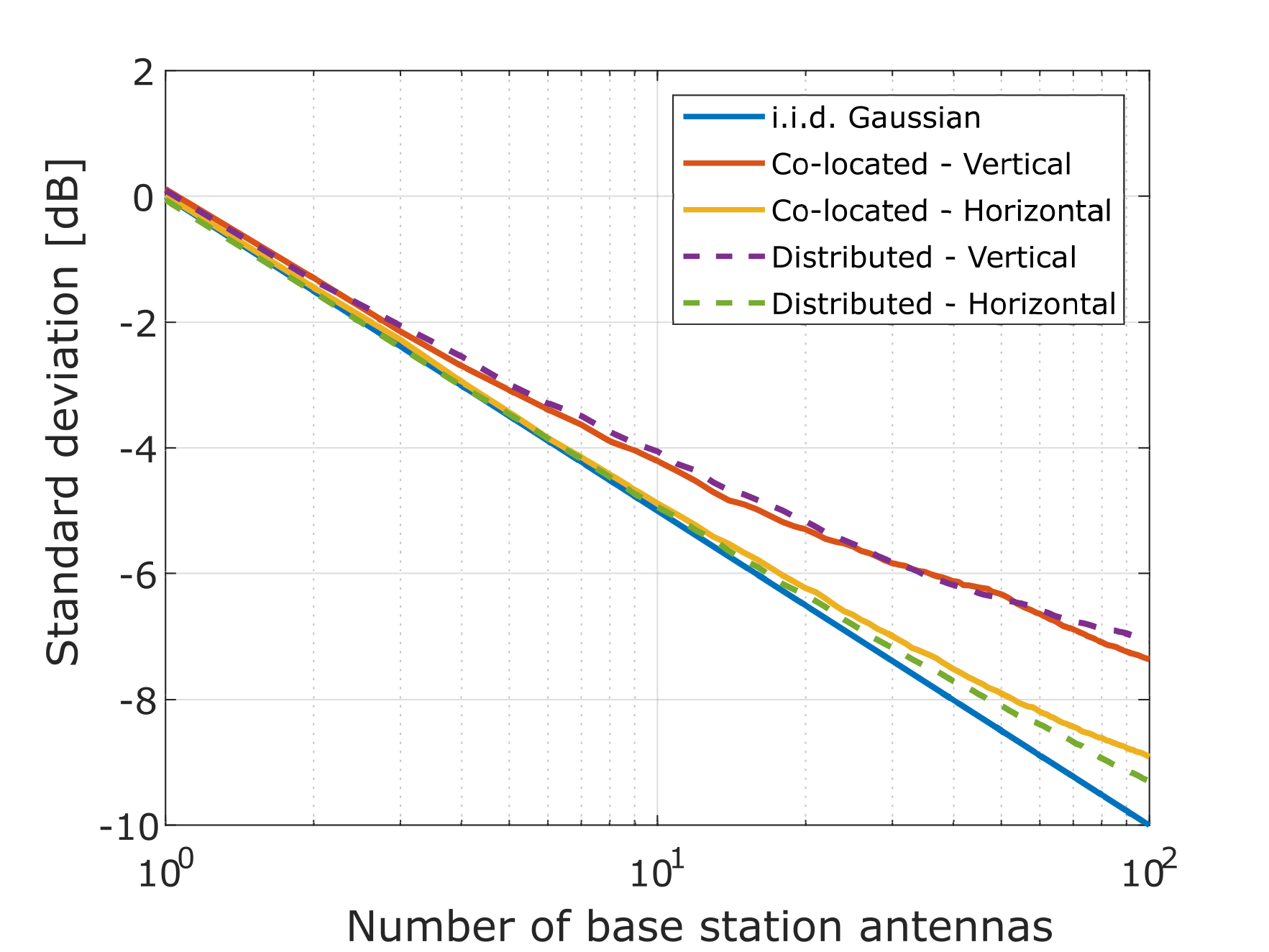}
    \caption{Channel hardening when scanning the left aisle for both arrays and UE orientations. The i.i.d. complex Gaussian channel is shown for reference.}
    \label{fig:chhard-left-curve}
    \vspace{-0.3in}
\end{figure}

A visualization of the array gain and channel hardening effect is presented in Fig.~\ref{fig:chhard-left} for the co-located array in the NLoS part of the first experiment. Here, the channel gain when combining all 100 antennas in comparison to the one antenna case can be seen as a function over time and frequency. When adding up the signals of the 100 antennas, the mean of the channel gain increases, that is the array gain. What also can be seen is that for the one antenna case, there are large variations over both time and frequency, while for the 100 antenna case the variations are just a few dB; this is the channel hardening effect. 


To quantify the experienced channel hardening, Fig.~\ref{fig:chhard-left-curve} shows the decrease of standard deviation as a function of the number of BS antennas. The standard deviation for the i.i.d. complex Gaussian is shown for reference. Furthermore, the standard deviation for both the UE with vertical and horizontal orientation are shown and a comparison of the co-located and distributed array. It can be seen that there is indeed a very prominent channel hardening effect as the standard deviation of channel gain decreases as the number of BS antennas increases. It can also be concluded that there are no major differences between the co-located and the distributed array; they behave similarly such that for both arrays, the horizontally oriented UE is experiencing more channel hardening. In general, the vertical UE orientation should be better at picking up signals reflected from machinery on the sides while the horizontal UE orientation ought to be better placed to pick up ceiling and floor reflections.

The measured standard deviations are following the theoretical curve quite closely, or with a small offset, for a smaller number of antennas. For a higher number of BS antennas, the curves start to diverge as the channel hardening effect saturates. However, two of the measured standard deviations in Fig.~\ref{fig:chhard-left-curve} follow the theoretical one very closely, showing that this is indeed a very rich scattering environment. The resulting channel hardening, measured as the difference between the end and starting points of each curve, is between 7.1~dB and 9.2~dB. This is not that far from the theoretical benchmark, which for 100~BS antennas is 10~dB. Comparing to a more general indoor environment, \cite{chhard_journal} measured in an indoor auditorium in LoS, reaching standard deviations down to -6.8~dB, although less than that in most cases. This indicates that in our indoor rich scattering factory environment, the channels between the UE and the different BS antennas are to a large extent independent, and hence a large portion of the channel hardening effect can be harvested, approaching the theoretical bound.

\vspace{-0.2in}

\section{Modeling the tails of the CDFs}

\begin{figure}[t]
    \centering
    \includegraphics[width=0.4\textwidth]{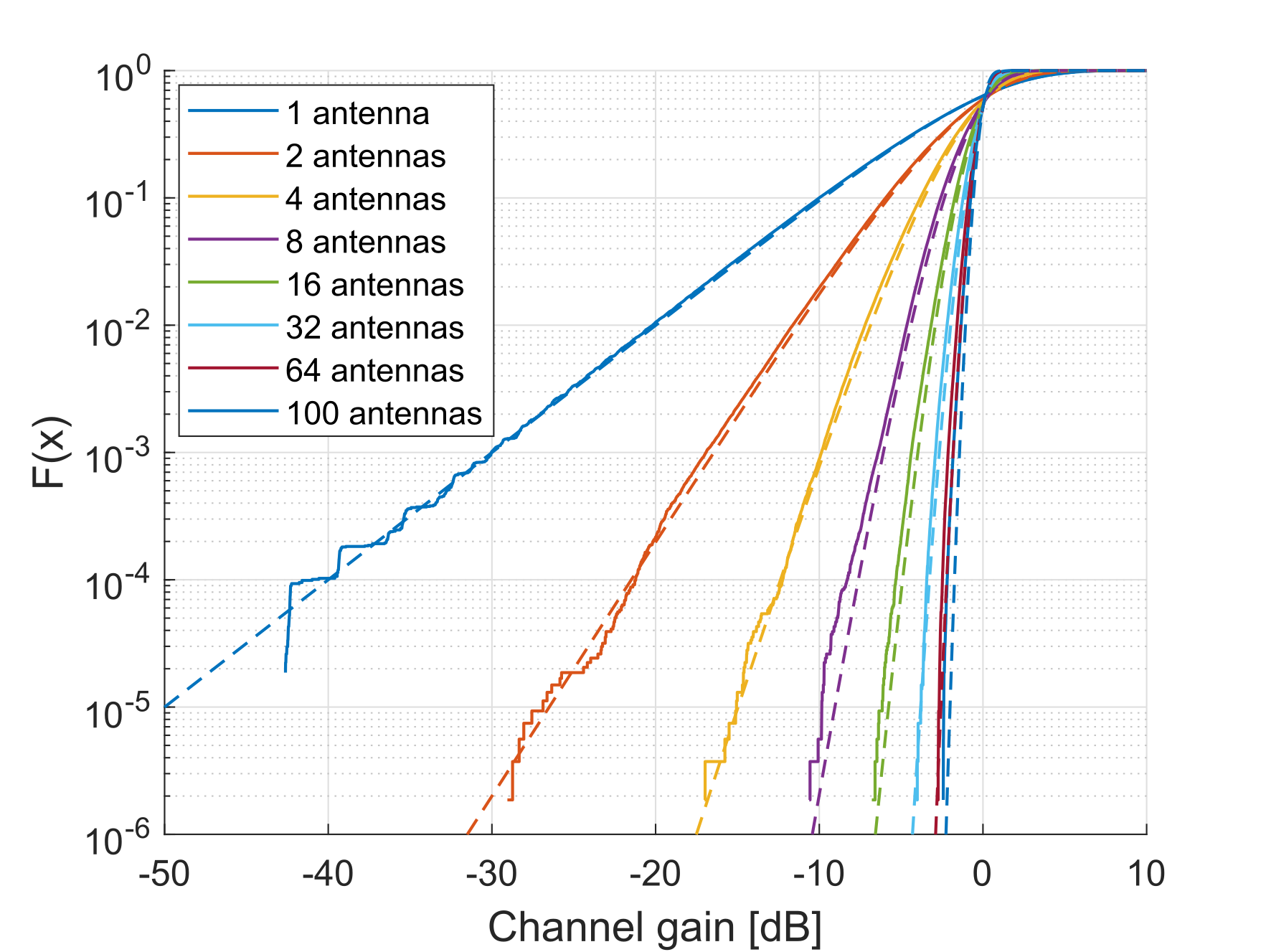}
    \caption{CDF of the channel gain when scanning the left aisle with horizontal UE orientation and the co-located array deployed. The gamma distribution ($\Gamma(M,1/M)$) is shown for reference.
    }
    \label{fig:cdf-left-co-log}
    \vspace{-0.2in}
\end{figure}

\begin{figure}[t]
    \centering
    \includegraphics[width=0.4\textwidth]{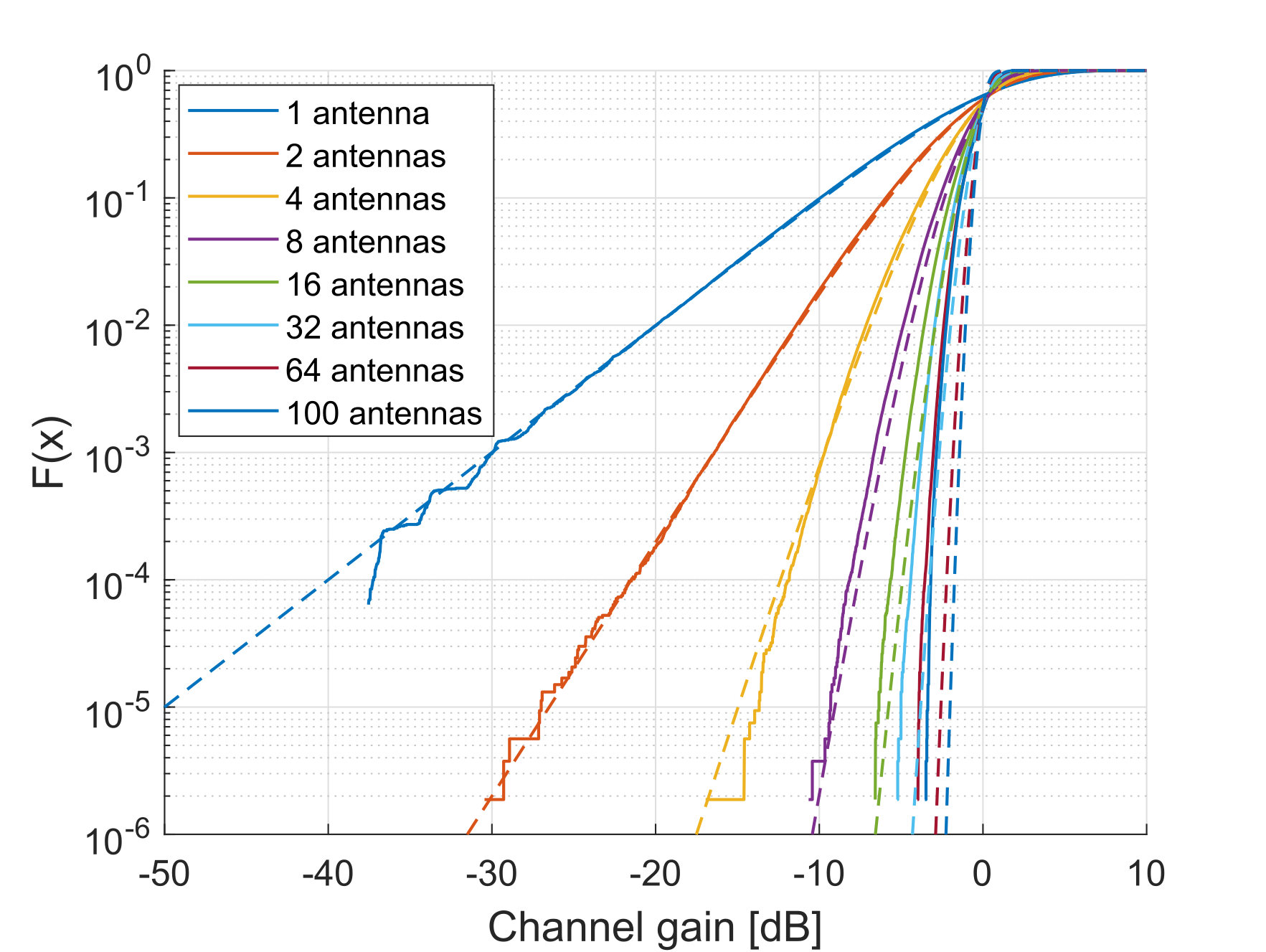}
    \caption{CDF of the channel gain when scanning the left aisle with vertical UE orientation and the distributed array deployed. The gamma distribution ($\Gamma(M,1/M)$) is shown for reference.
    }
    \label{fig:cdf-left-dist-log}
    \vspace{-0.2in}
\end{figure}

\begin{figure}[t]
    \centering
    \includegraphics[width=0.4\textwidth]{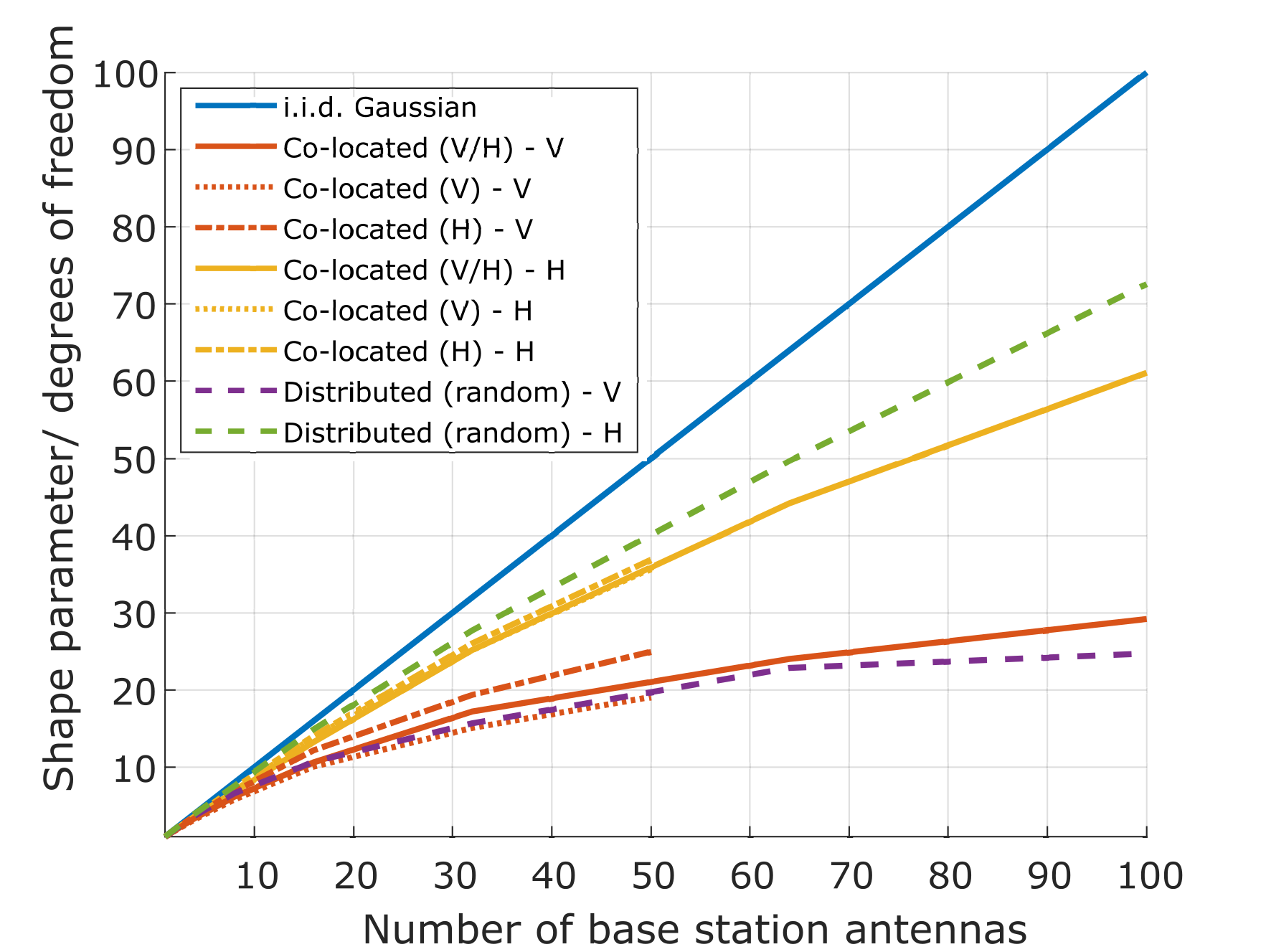}
    \caption{The estimated shape parameter $\hat{M}$ for both the co-located and the distributed array when scanning the left aisle with  the vertical or horizontal UE orientation. The i.i.d. complex Gaussian channel is shown for reference.}
    \label{fig:shape}
    \vspace{-0.2in}
\end{figure}

\begin{figure}[t]
    \centering
    \includegraphics[width=0.4\textwidth]{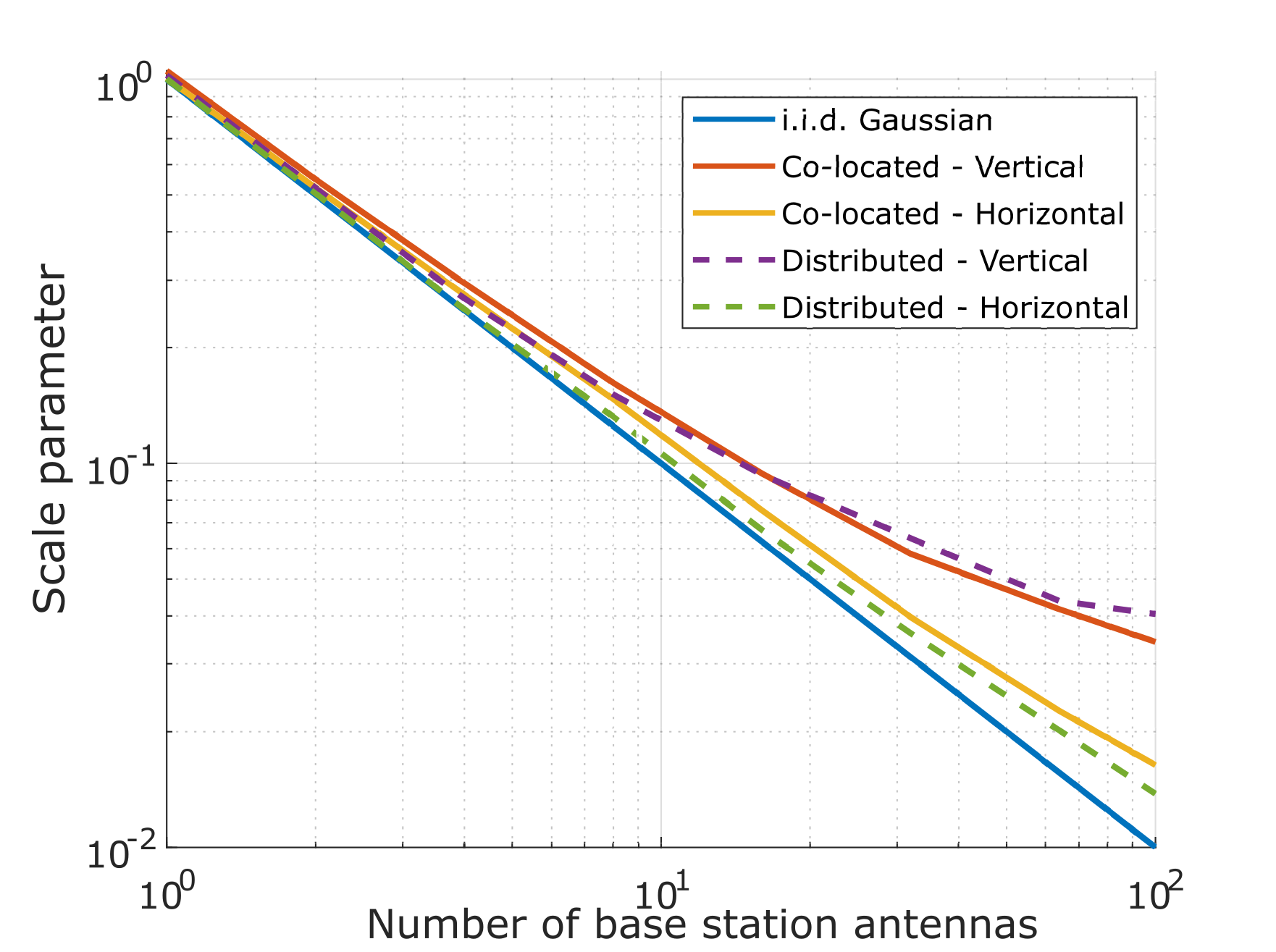}
    \caption{The estimated scale parameter $\frac{1}{\hat{M}}$ for both the co-located and the distributed array when scanning the left aisle with the vertical or horizontal polarization. The i.i.d. complex Gaussian channel is shown for reference.}
    \label{fig:scale}
    \vspace{-0.3in}
\end{figure}

Here, the CDFs are modeled by assuming that the underlying channel coefficients $h$ are i.i.d. complex Gaussian, leading to Rayleigh distributed $|h|$ and the channel gain $|h|^2$ resulting in an exponential distribution. Summing up the power for the $M$ BS antennas as done with maximum ratio combining/transmission (MRC/MRT), a gamma distribution $\Gamma(M,1/M)$ is expected. The shape parameter, here ideally translated to $M$, can be seen as a measure of the possible degrees of freedom (DoF), which in case of fully independent channel coefficients would be the number of BS antennas.

Visualizing the same measured channel gains but in another analysis, Fig.~\ref{fig:cdf-left-co-log} and \ref{fig:cdf-left-dist-log} show the CDFs of the channel gain for $M=[1, 2, 4, 8, 16, 32, 64, 100]$, for the co-located and the distributed configuration. The array gain, as seen in Fig.~\ref{fig:chhard-left}, can not be seen here, since the gains are normalized by $M$, meaning that the output power from the BS could be reduced with a factor $M$. The channel hardening can however be clearly seen as the CDFs get steeper with more antennas. The major improvements happen at the beginning and as the antenna set gets larger, the difference is not as significant, although small improvements are still to be harvested, in line with Fig.~\ref{fig:chhard-left-curve}.

As a reference, the gamma distributions $\Gamma(M,1/M)$ with the shape parameter $M$ and scale parameter $1/M$ is shown with dashed lines. It can be observed that for a smaller subset of antennas, i.e. up to about 32 BS antennas, both the scale and shape parameter correspond well to the measurements. However, after that, there is a visible difference, indicating that one additional DoF is not gained by adding one more BS antenna and hence, all channels are not independent of each other. This observation can be made for both the co-located and the distributed array. It can also be observed that the slope for the measured curves in general are matching while the offset of the slope at e.g. $10^{-5}$ for $M>2$ is between $0.02-0.80$ for the co-located array with horizontal UE orientation and $0.52-1.37$ in the distributed case with vertical UE orientation.

Elaborating on the previous observation, the shape and scale parameter are shown in Fig.~\ref{fig:shape} and \ref{fig:scale} for the co-located and the distributed setup for both UE orientations. The case when only choosing the vertically or horizontally polarized BS antennas is also included. In the case where the underlying channel coefficient would be i.i.d. complex Gaussian, the shape parameter would increase by 1 when adding 1 antenna, i.e. adding 1 DoF for each new BS antenna. It can be observed that this is indeed not entirely the case, although not too far away for one of the two possible UE orientations. These results can easily be related to the channel hardening in Fig.~\ref{fig:chhard-left-curve}, serving as the link indicating how many independent channels there actually are for a certain subset of antennas. As an example. for the channel hardening curve in Fig.~\ref{fig:chhard-left-curve}, the distributed setup with horizontal UE orientation ends up at the same point as the i.i.d. complex Gaussian for $73$~antennas. This can then be seen in the shape parameter in Fig.~\ref{fig:shape}. Interesting to note is also that the slopes when only choosing one polarization at the BS are similar to when alternating, possibly due to the fact that most paths may be reflected.
The scale parameter, which can be translated to the standard deviation, is shown in Fig.~\ref{fig:scale} and can also be used as a measure to quantify the channel hardening effect; here seen as the decreasing slope as the number of BS antennas increases, which is corresponding to the slopes in Fig.~\ref{fig:chhard-left-curve}. 
\vspace{-0.1in}






\section{Fading margin}

In \cite{fading_margin}, the link between channel hardening and fading margin is elaborated on and a measurement-based definition of the fading margin is introduced by using the empirical CDF (ECDF). The definition relates to the median, as dependent on the number of antennas $M$ and with probability $p$, and is given as

\vspace{-0.1in}

\begin{equation}
    F_M(p)=10\text{log}_{10}(\frac{Q(0.5)}{Q(p)})
    \label{fading_margin}
\end{equation}

\noindent where $Q$ is the quantile/inverse CDF function. Naturally, the more channel hardening, the steeper the CDF, and the smaller fading margin is required. 


\begin{figure}[t]
    \centering
    \includegraphics[width=0.4\textwidth]{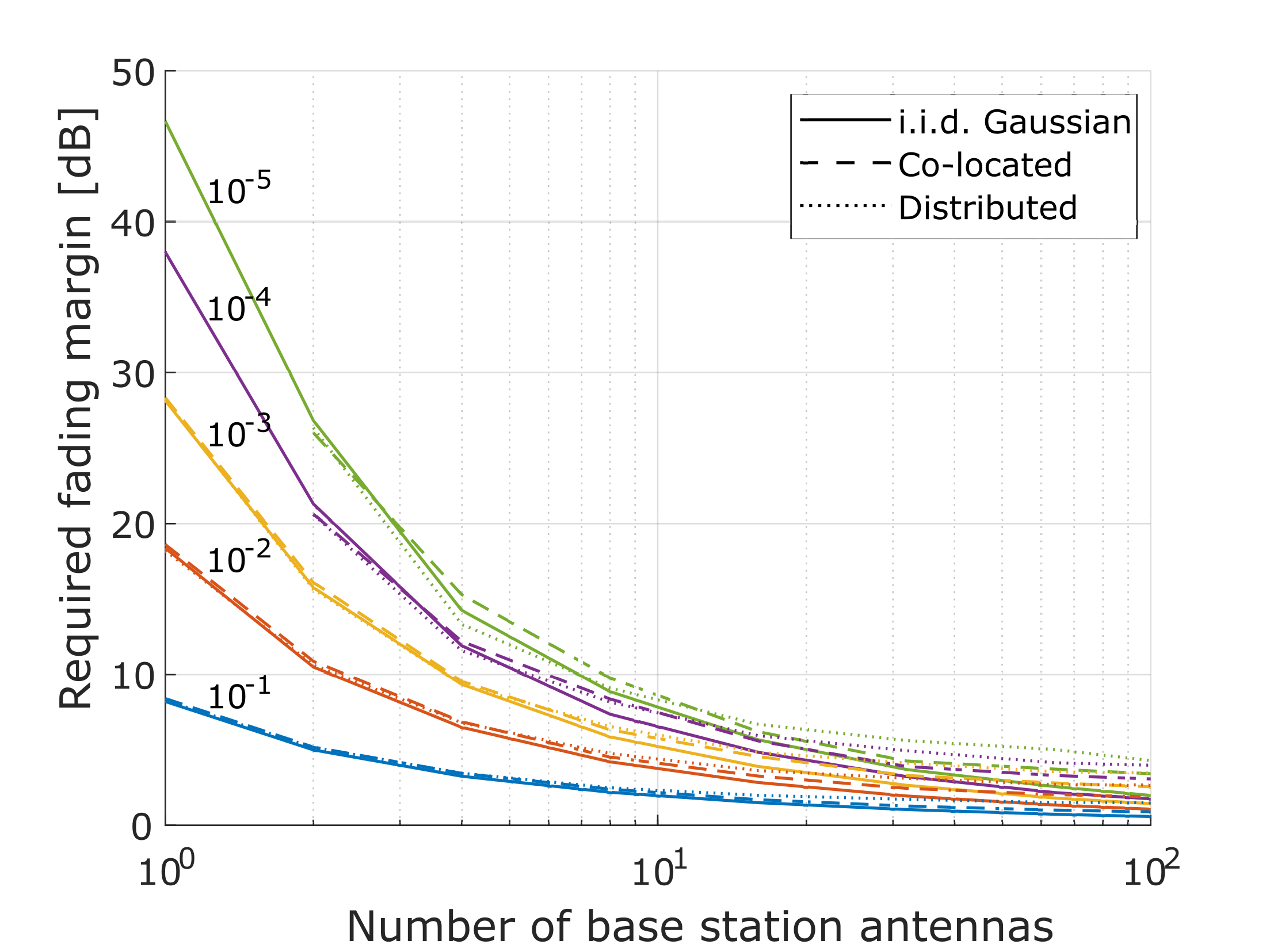}
    \caption{The required fading margin for different outage thresholds for the co-located and the distributed array.
    }
    \label{fig:fm-left}
    \vspace{-0.2in}
\end{figure}

In Fig.~\ref{fig:fm-left}, the required fading margin, as given in (\ref{fading_margin}) to achieve different outages is shown as a function of the number of BS antennas for the left aisle for the co-located array with horizontal UE orientation and the distributed array with vertical UE orientation. Outages between $10^{-5}-10^{-1}$ and antenna subsets from $[1, 2, 4, 8, 16, 32, 64, 100]$ are shown, as this is what can be used based on the data, as seen in Figs.~\ref{fig:cdf-left-co-log} and \ref{fig:cdf-left-dist-log}. The i.i.d. complex Gaussian channel is shown for reference. What is seen is steadily decreasing required fading margins as the number of BS antennas increases. The co-located and the distributed array are following each other very well. Starting from one antenna, the required fading margins decrease quite rapidly, meaning that there is a lot to gain in terms of fading margin by adding just a few, while the gain in terms of decreased fading margin is not that big when compared to the larger subsets of antennas. 
It can be noted that at $100$ antennas and an outage threshold of $10^{-5}$, the required fading margin is not exceeding $4.3$~dB.

\vspace{-0.1in}

\section{Shadowing}

\begin{figure}[t]
    \centering
    \includegraphics[width=0.4\textwidth]{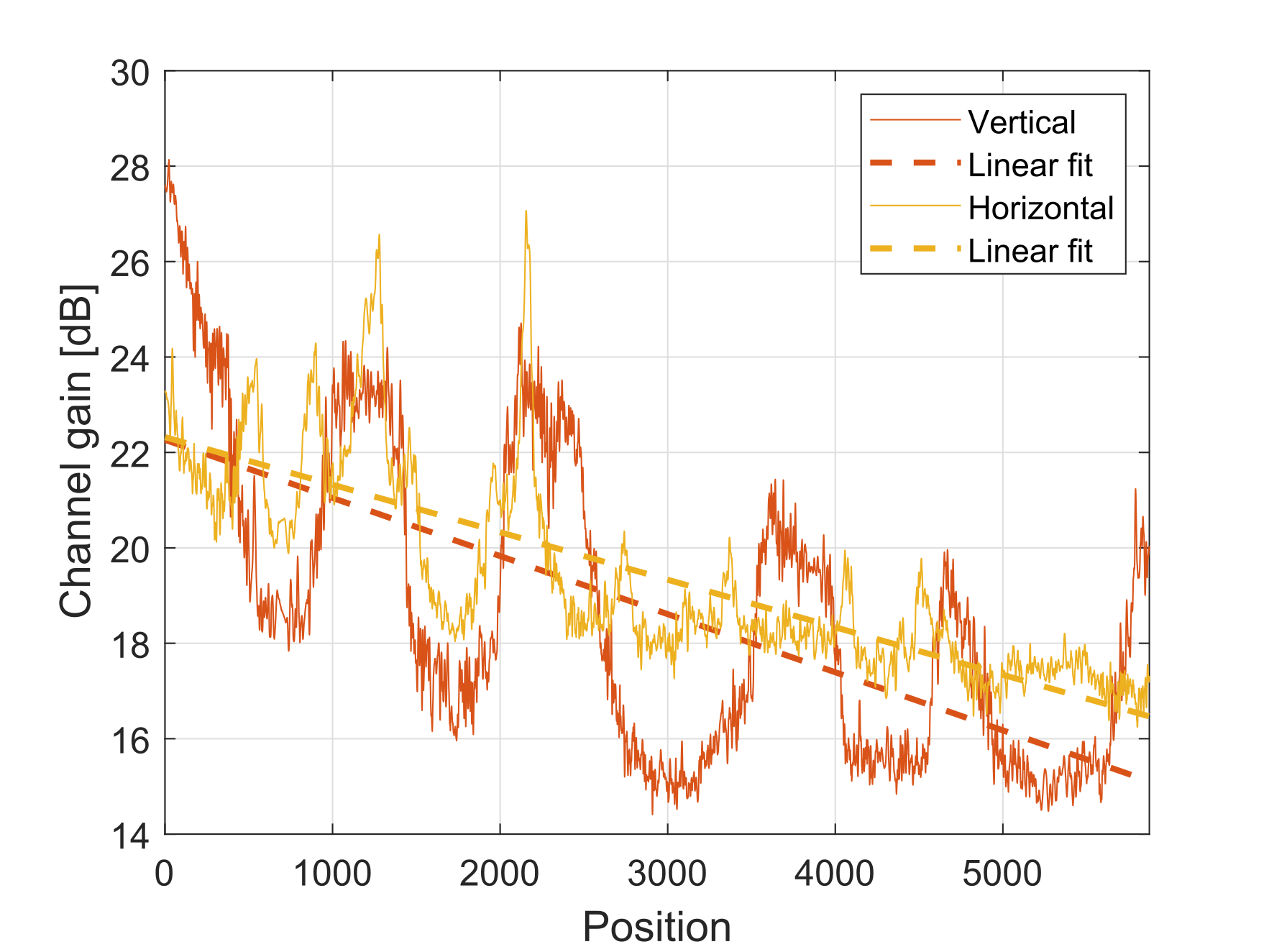}
    \caption{Channel gain over time for the co-located array and both UE orientations with the linear regression line.}
    \label{fig:co-pl}
    \vspace{-0.2in}
\end{figure}

\begin{figure}[t]
    \centering
    \includegraphics[width=0.4\textwidth]{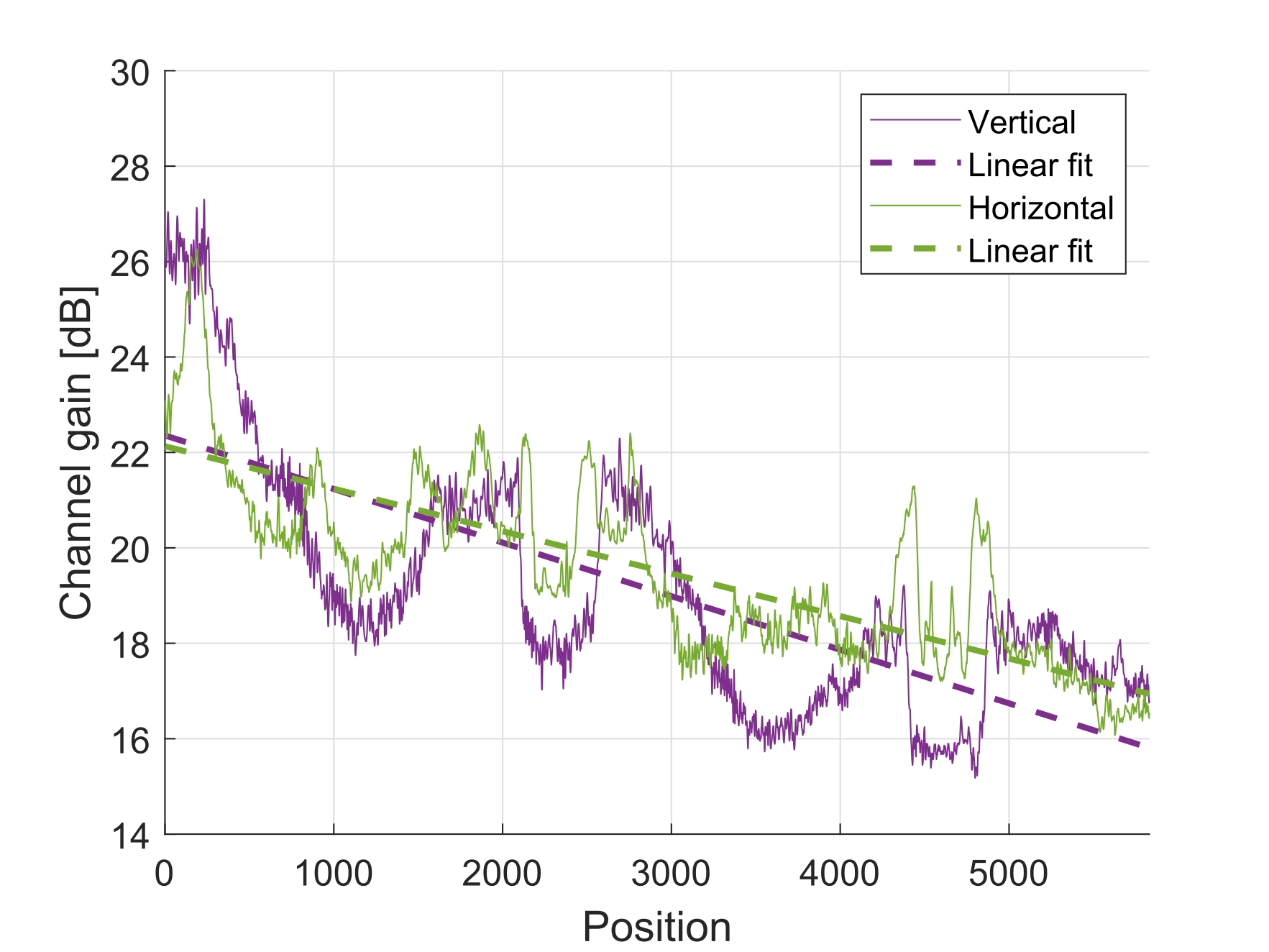}
    \caption{Channel gain over time for the distributed array and both UE orientations with the linear regression line.}
    \label{fig:dist-pl}
    \vspace{-0.2in}
\end{figure}

\begin{figure}[t]
    \centering
    \includegraphics[width=0.4\textwidth]{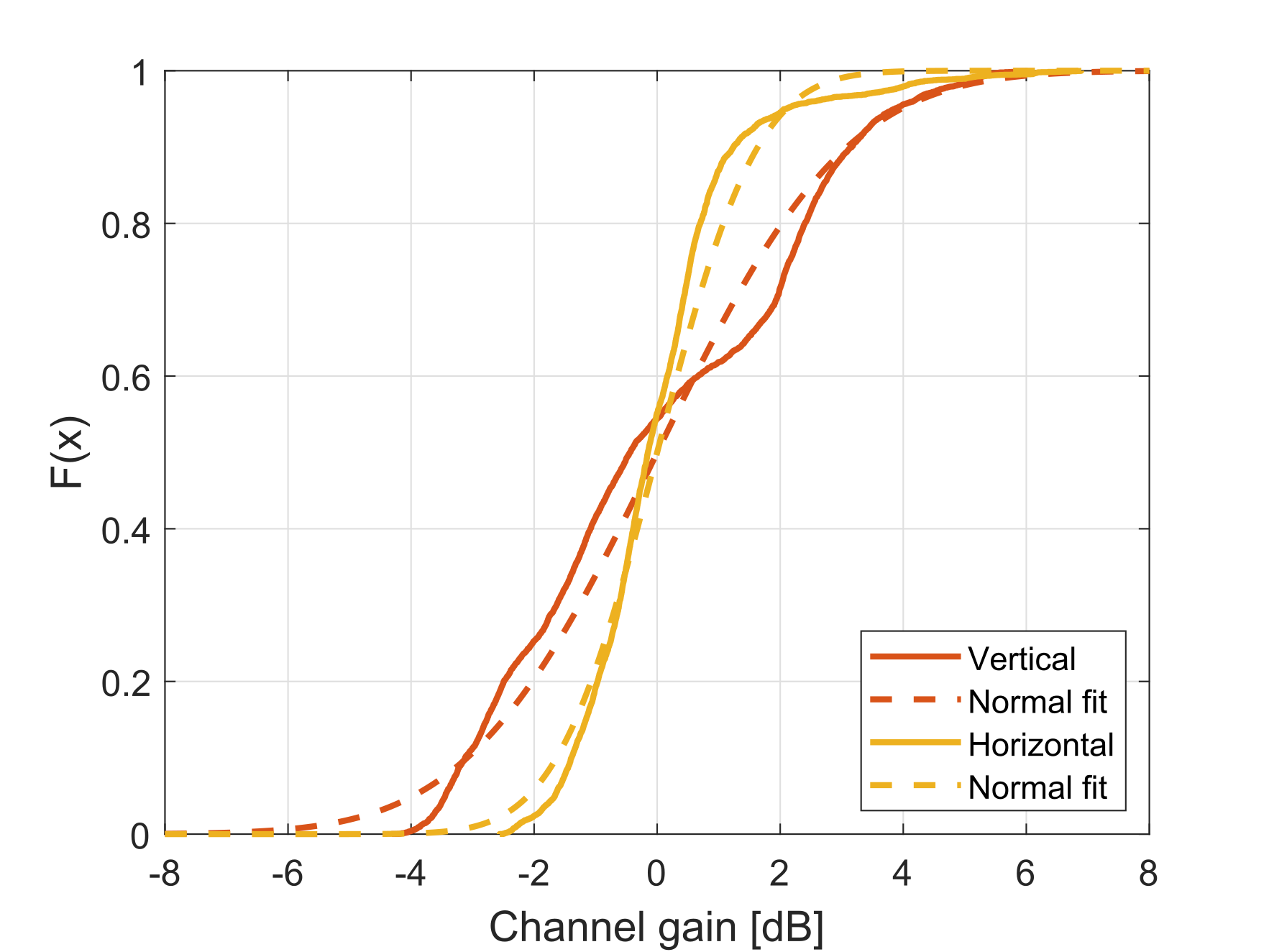}
    \caption{CDF of the shadowing for the co-located array and both UE orientations and the fitted normal distribution.}
    \label{fig:co-shadow-cdf}
    \vspace{-0.2in}
\end{figure}

\begin{figure}[h]
    \centering
    \includegraphics[width=0.4\textwidth]{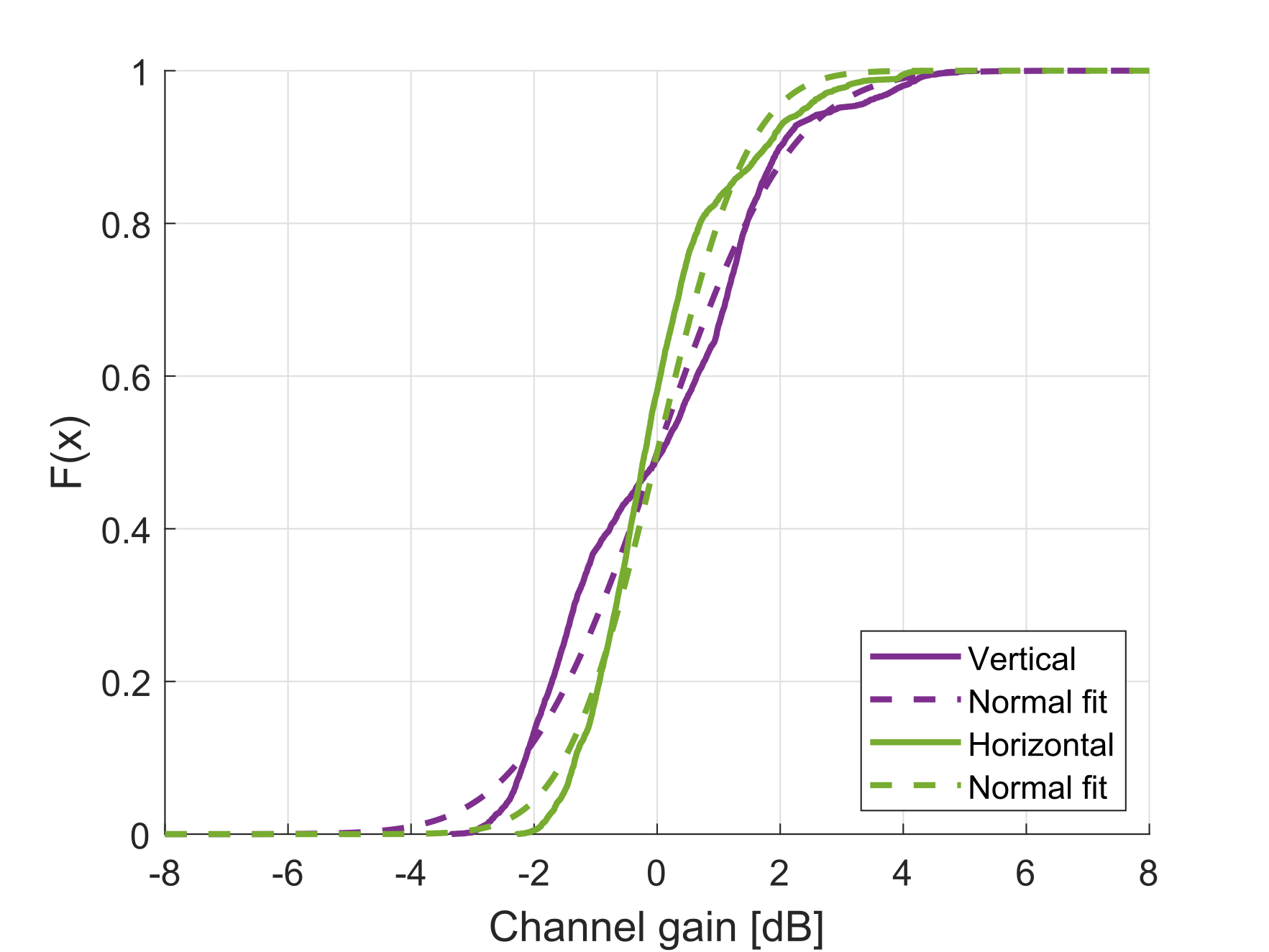}
    \caption{CDF of the shadowing for the distributed array and both UE orientations and the fitted normal distribution.}
    \label{fig:dist-shadow-cdf}
    \vspace{-0.2in}
\end{figure}

Changing the focus towards large-scale fading and shadowing effects, the analysis continues with the results from the second experiment, Figs.~\ref{fig:co-pl} and \ref{fig:dist-pl} show the channel gain for different positions for each BS array, as averaged over frequency and summed for all BS antennas. The effect is as expected a high channel gain when being in LoS in the corridor and a lower channel gain when being shadowed in the aisles. Here the first four to five aisles can be seen for both the vertical and horizontal UE orientation, which show similar behaviour globally but they also show differences when analyzing the details. For example, the horizontal UE does not experience as severe dips when being shadowed, potentially due to being better at picking up reflections from the ceiling. The channel gain is also decreasing from left to right, which is an effect of the path-loss when increasing the distance. However, note that the x-axis is not directly translatable to distance as it is not strictly increasing with the position/measurement time, as seen in Fig.~\ref{fig:map}. 

To investigate the shadowing, the channel gains are fitted to a line ($y=kx+m$) by linear regression in a least-square sense and the values for $k$ and $m$ is shown in Table~\ref{tab:shadowing}. The shadowing is analysed as the variations around this line and Figs.~\ref{fig:co-shadow-cdf} and \ref{fig:dist-shadow-cdf} show the CDF of channel gain (on a dB-scale) for this shadowing for the two array configurations. For each array, the vertical and horizontal UE are shown with their respective fit to a normal distribution, resulting in a log-normal distribution, with the standard deviation given in Table~\ref{tab:shadowing}.

\begin{table}[h]
    \vspace{-0.1in}
    \centering
    \begin{tabular}{cc|ccc}
        & & $\hat{k}$ & $\hat{m}$ & $\hat{\sigma}$ \\
        \hline
         \multirow{2}{*}{Co-located} & V & -0.0012 & 22.26 & 2.41 \\
         & H & -0.0010 & 22.32 & 1.28\\
         \hline
         \multirow{2}{*}{Distributed} & V & -0.0011 & 22.36 & 1.72\\
         & H & -0.0009 & 22.13 & 1.17\\
    \end{tabular}
    \caption{Estimated parameters as shown in Figs.~\ref{fig:co-pl}-\ref{fig:dist-shadow-cdf}.}
    \label{tab:shadowing}
    \vspace{-0.1in}
\end{table}

In Table~\ref{tab:shadowing}, the largest standard deviation is found to be with the co-located array, serving the vertical UE, and the smallest variance is with the distributed array serving the horizontal UE. The other two UEs are somewhere in between. For the four cases seen in Figs.~\ref{fig:co-pl}-\ref{fig:dist-shadow-cdf}, the normal fit overestimates the tail; the measured shadowing is less severe than this when it comes to the more extremes. For the measured channels, the shadowing variation is between $-4$ and $6$~dB for the co-located array and $-3$ and $5$~dB for the distributed array. These results indicate that the worst cases of shadowing can be avoided and the span can be reduced with a distributed array.
Especially since, again reminding of the fact that the co-located array is pointing in the direction of the experiment while the distributed is somewhat random, the UE movement is unknown and therefore the best placement of a co-located array would vary while in the distributed case, some of the antennas should always have a more beneficial placement.


\vspace{-0.1in}

\section{Conclusions}

We have investigated the channel characteristics of massive MIMO, deployed in a real operating factory. The environment provides blocking effects but also extremely rich scattering, leading to the fact that signals are coming from almost every angle and that there are no dominant directions. This results in channel responses being not far from coinciding with the often theoretically assumed i.i.d. complex Gaussian channel.

With massive MIMO, the channel becomes more reliable through the array gain and channel hardening. The channel hardening can also be viewed as the CDF of channel gains becomes steeper, here compared to a gamma distribution, whose parameters have been estimated, giving a slightly different view of this effect and a relation to the number of DoF. The channel hardening effect results in that very small fading margins are required and that has also been quantified. With $100$~antennas the required fading margin does not exceed $4.3$~dB for any of the two array configurations.

Both a co-located and a fully distributed array have been deployed, aiming at further extracting spatial diversity, and comparisons have been made. The conclusion from this analysis is that the small-scale fading statistics are similar for the two arrays, however, the worst cases in terms of large-scale power variations are avoided and the span of experienced shadowing is reduced with a distributed setup, being beneficial in heavily shadowed environments and unknown UE movement patterns. 

Having an almost i.i.d. complex Gaussian channel will impact the whole system architecture, suggesting one to rethink how the channel relates to both the hardware design and the upper layers. From a system perspective, the channel coding can be done with short block lengths and short packets. A very limited number of re-transmissions should also be possible, enabling ultra-low latency.
To summarize; xURLLC can be realizable in reflective and heavily shadowed industrial environments with large distributed antenna arrays, but co-located arrays can also be sufficient to achieve this goal.


\section*{Acknowledgment}
The authors would like to thank Harsh Tataria, Steffen Malkowsky and Anders J Johansson for the help during the measurement campaign and the Bosch staff at the semiconductor factory in Reutlingen for enabling this work. 


\bibliographystyle{IEEEtran}

\bibliography{current_bib}

\vfill

\end{document}